\documentclass{aa}
\usepackage{graphicx}
\usepackage{natbib}
\begin{document}
%

   \title{First determination of the dynamical mass of a binary L dwarf \thanks{Based on observations obtained with the the NASA/ESA \emph{Hubble Space Telescope}, the ESO \emph{Very Large Telescope} (program 70.D-0773), and the \emph{W. M. Keck Observatory}.}}


   \author{H. Bouy \inst{1,2,4}, G. Duch\^ene \inst{4,6}, R. K\"ohler \inst{3}, W. Brandner \inst{3}, J. Bouvier \inst{4}, E.~L. Mart\'\i n \inst{5}, A. Ghez \inst{6}, X. Delfosse \inst{4}, T. Forveille \inst{4,10}, F. Allard \inst{7}, I. Baraffe \inst{7}, G. Basri \inst{8},  L. Close \inst{9}, C.~E. McCabe \inst{6}}

   \offprints{H. Bouy}

   \institute{Max Planck f\"ur Extraterrestrische Physik,  Giessenbachstra\ss e, D-85748 Garching bei M\"unchen, Germany\\
              \email{hbouy@mpe.mpg.de}
	      \and European Southern Observatory, Karl Schwarzschildstra\ss e 2, D-85748 Garching bei M\"unchen, Germany\\
              \and Max-Planck Institut f\"ur Astronomie, K\"onigstuhl 17, D-69117 Heidelberg, Germany\\
              \email{brandner@mpia.de, koehler@mpia.de} 
              \and Laboratoire d'Astrophysique de l'Observatoire de Grenoble, 414 rue de la piscine, F-38400 Saint Martin d'H\`ere, France\\
	     \email{Jerome.Bouvier@obs.ujf-grenoble.fr, Gaspard.Duchene@obs.ujf-grenoble.fr}\\
	     \email{Xavier.Delfosse@obs.ujf-grenoble.fr}
              \and Instituto de Astrofisica de Canarias, 38200 La Laguna, Spain\\
              \email{ege@ll.iac.es}
              \and Division of Astronomy \& Astrophysics, University of California in Los Angeles, Los Angeles, CA90095-1562, U.S.A\\
              \email{ghez@astro.ucla.edu, mccabe@astro.ucla.edu}
              \and Centre de Recherche Astronomique de Lyon (UML 5574), Ecole Normale Sup\'erieure, 69364 Lyon Cedex 07, France\\
              \email{fallard@ens-lyon.fr, ibaraffe@ens-lyon.fr}
	      \and University of California at Berkeley, Astronomy Department, MC 3411 Berkeley, CA 94720, U.S.A\\
	      \email{basri@astro.berkeley.edu}
	      \and Steward Observatory, University of Arizona, 933 N. Cherry Ave., Tucson, AZ 85721-0065, U.S.A\\
	      \email{lclose@as.arizona.edu}
	      \and Canada-France-Hawaii Telescope Corporation, PO Box 1597, Kamuela Hi-96743, USA\\
	      \email{forveille@cfht.hawaii.edu}
      }

   \date{Received 29/03/2004; accepted 05/05/2004}

   \abstract{We present here the results of astrometric, photometric and spectroscopic observations leading to the determination of the orbit and dynamical masses of the binary L dwarf 2MASSW~J0746425+2000321. High angular resolution observations spread over almost 4 years and obtained with the \emph{Hubble Space Telescope} (HST), the ESO \emph{Very Large Telescope} (VLT), and a the \emph{W.~M. Keck Observatory} (Keck) allow us to cover $\sim$36\% of the period, corresponding to 60\% of the orbit, and, for the first time, to derive a precise estimate of the total and individual masses of such a late-type object. We find an orbital period of  3850.9$^{+904}_{-767}$~days. The corresponding total mass is 0.146$^{+0.016}_{-0.006}$~M$_{\sun}$, with uncertainties depending on the distance. Spatially resolved low resolution optical (550--1025~nm) spectra have been obtained with HST/STIS, allowing us to measure the spectral types of the two components (L0$\pm$0.5 for the primary and L1.5$\pm$0.5 for the secondary). We also present precise photometry of the individual components measured on the high angular resolution images obtained with HST/ACS and WFPC2 (visible), VLT/NACO (J, H and K$_{S}$ bands) and Keck~I (K$_{S}$ band). These spectral and photometric measurements enable us to estimate their effective temperatures and mass ratio, and to place the object accurately in a H--R diagram. The binary system is most likely formed by a primary with a mass of 0.085$\pm$0.010~M$_{\sun}$ and a secondary with a mass of 0.066$\pm$0.006~M$_{\sun}$, thus clearly substellar, for an age of approximately 300$\pm$150~Myr. H$\alpha$ variability indicates chromospheric and/or magnetic activity.
 
    \keywords{- stars: very low-mass, brown dwarfs - star: \object{2MASSW~J0746425+2000321} - Binary: orbit, dynamical mass}
}

\authorrunning{Bouy et al.}
\titlerunning{Dynamical mass of a binary L dwarf}

\maketitle

\markboth{Dynamical mass of a binary L dwarf}{Bouy et al.}

\section{Introduction}

With spectral properties intermediate between those of giant planets and late-type stars, ultra-cool and brown dwarfs have opened a new chapter in the study of atmospheric physics and chemistry. One of the ultimate goals of a theory of very low mass and sub-stellar objects is an accurate determination of the mass of an object based on spectroscopic characteristics and luminosity. The degeneracy in the mass-luminosity (age-temperature) relation for ultra-cool dwarfs makes it difficult to pin down their physical properties. Luminosities and effective temperatures of ultra-cool dwarfs are function of both age and mass \citep{1997ApJ...491..856B} so that an older, slightly more massive ultra-cool dwarf can exhibit the same effective temperature as a younger, less massive one. Therefore, dynamical masses, which are model-independent, are highly required in order to calibrate the mass-luminosity relation. The study of binary ultra-cool dwarfs offers several advantages for such a study. Both components of the binary are expected to be coeval, thus removing part of the above mentioned degeneracy in the mass-luminosity (age-temperature) relation. Only very few observational constrains on the masses of this class of objects are available nowadays, and we present here the first measurement for field L-dwarfs, at the stellar/sub-stellar transition. The only one similar measurement available up to now concerned the M8.5/M9.0 brown dwarf binary \object{GJ~569Bab}, presented by \citet{2001ApJ...560..390L}. Although the age of 2MASSW~J0746425+2000321 is not known indenpendently from any models yet, these observations give promising results as a first step toward the calibration of the models.

In this paper, we will present the results of high angular resolution imaging and low resolution spectroscopic observations of a binary L dwarf, leading to the determination of its orbital parameters, and total and individual masses. In section \ref{target}, we will present 2MASSW~J0746425+2000321, in section \ref{obs} we will describe the observations and the processing of the data, in section \ref{analysis} we will explain the calculation of the orbital parameters and in section \ref{discussion} we will present the analysis of the individual spectra and luminosities and discuss the results.

\section{2MASSW~J0746425+2000321 \label{target}}
\object{2MASSW~J0746425+2000321} has been observed and reported in several catalogues and articles. It has been identified as a L0.5 dwarf by \citet{2000AJ....120..447K}, and suggested to be a binary by \citet{2000AJ....119..369R} based on its position in a colour-magnitude diagram. It has been resolved as a multiple system by \citet{2001AJ....121..489R} with a separation of 0\farcs22 and a position angle (P.A) of 15\degr, a measurement later corrected by \citet{2003AJ....126.1526B} to 0\farcs219$\pm$0\farcs003 and P.A=168\fdg8$\pm$0\fdg3. 2MASSW~J0746425+2000321 has been reported in several surveys, such as USNO-B \citep{2003AJ....125..984M}, GSC2.2, and 2MASS \citep{2MASS}. Table \ref{catalogues} gives an overview of the astrometric and photometric properties of 2MASSW~J0746425+2000321 as reported in these catalogues. \citet{2002AJ....124.1170D} and the USNO-B.1 catalogue both report a measurement of the proper motion of this objects, with $\mu_{\alpha}$=-370$\pm$4~mas yr$^{-1}$ and $\mu_{\delta}$=-42$\pm$4~mas yr$^{-1}$ (USNO-B.1) and $\mu_{\alpha}$=-374$\pm$0.3~mas yr$^{-1}$ and $\mu_{\delta}$=-58$\pm$0.3~mas yr$^{-1}$ \citep{2002AJ....124.1170D}. These measurements confirm that 2MASSW~J0746425+2000321 is a common proper motion pair. Such a proper motion indeed implies a motion of $\sim$1\farcs5 during the 4 years we made the follow up observations, whereas the separation between the two components varied only of $\sim$0\farcs1. Using high resolution spectra obtained at Keck, \citet{2002AJ....124..519R} measured a rotational velocity of 24~km/s. Using VLT/UVES high resolution spectra, \citet[accepted for A\&A,][]{Bailer-Jones} measured a rotational velocity ranging between 25.6$\le v $\,sin$i \le$30.6 km/s, corresponding to a period between 1.73$\le$T$\le$3.71 hours. Both \citet{2002MNRAS.335.1158C} and \citet{2002ApJ...577..433G} report photometric variability, which they attribute mainly to the formation of clouds in the upper layers of the atmospheres. \citet{2002AJ....124.1170D} measured its distance using trigonometric parallax at 12.21$\pm$0.05~pc.

\section{Observation and data processing \label{obs}}

Table \ref{obs_log} gives a log of all the observations we use in this study.

\subsection{High Angular Resolution imaging with HST/ACS and STIS}
\begin{figure*}
   \centering
   \includegraphics[width=\textwidth]{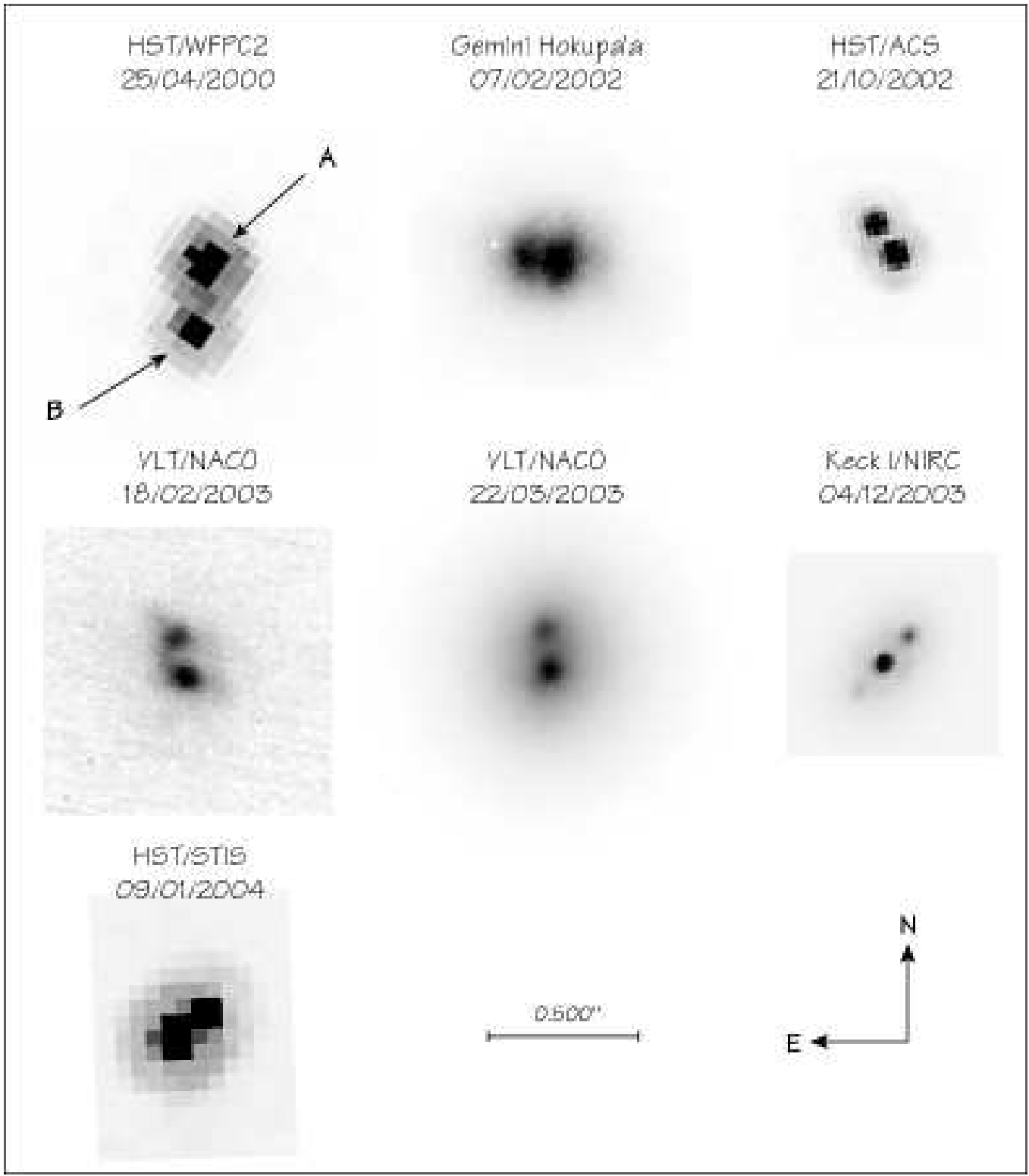}
   \caption{Images of 2MASSW~J0746425+2000321A and B obtained at different epochs with HST, Gemini, NACO and Keck I. The scale and the orientation are the same for all images, and indicated on the figure. \label{images}}
   \end{figure*}

High angular resolution images have been obtained with the HST Advanced Camera for Survey \citep[ACS,][]{ACS_INSTR_HANDBOOK} and Space Telescope Imaging Spectrograph \citep[STIS,][]{STIS_INSTR_HANDBOOK}. We observed 2MASSW~J0746425+2000321 using the ACS and its High Resolution Channel (HRC) in three different optical filters (F625W, F775W and F850LP), and STIS in the long-pass filter. 2MASSW~J0746425+2000321 is clearly resolved on both sets of data (see Figure \ref{images}), and we were able to get precise astrometric and photometric measurements. The data have been analyzed using a custom-made program performing PSF fitting to compute the precise separation, position angle and flux ratios of the multiple system. This program is identical to that used by \citet{2003AJ....126.1526B}, adapted for ACS/HRC and STIS, and is fully described in the mentioned paper. A brief summary is given here for completeness: the PSF fitting routine builds a model binary using ten different PSF stars coming from different ACS/HRC images. A cross-correlation between the model and the binary system yields the best values for the free parameters: separation, position angle and fluxes. The uncertainties and limitations of this technique are discussed in detail in \citet{2003AJ....126.1526B}. The program has been slightly improved since \citet{2003AJ....126.1526B} presented it in their paper, and the systematic errors and uncertainties on the difference of magnitude are now much better understood and corrected \citep{bouy2004}. While the original program was using only 5 free parameters (the coordinates of the primary, the coordinates of the secondary, and the flux ratio between the two components), the new version uses 6 free parameters (the 4 coordinates, the flux ratio between the primary and the PSF star, the flux ratio between the secondary and the PSF star), which allows to decrease considerably the systematic errors and uncertainties. 

\subsection{High Angular Resolution imaging with VLT/NACO}
We also obtained high angular resolution images using the ground based facilities offered by ESO on Cerro Paranal on 2003 February 18th and 2003 March 22nd. The VLT on \emph{Yepun} uses NACO, an adaptive optics platform \citep{2003SPIE.4839..140R, 2003SPIE.4841..944L, 2002Msngr.107....1B} to achieve diffraction limited images. NACO offers the possibility to use an infrared wavefront sensor, and is therefore ideally suited for the study of ultra-cool and red objects. Its CONICA array offers a 0\farcs01326$\pm$0\farcs001 pixel-scale that provides critical Nyquist sampling of the diffraction limited images of the telescope at these wavelengths. Its absolute orientation is known to  within $\sim$1\degr.

The atmospheric conditions during the observations were good (respectively $\lambda/r_{0}$=0\farcs62 and airmass=1.5, and $\lambda/r_{0}$=0\farcs67 and airmass=1.4), and very sharp images in K$_{S}$ (first observation) and J, H and K$_{S}$ (second observation) were obtained with strehl ratios of $S_{r}(K_{S})\sim$30\% (first obs.) and $S_{r}(J)\sim$13\%  ,$S_{r}(H)\sim$27\%, and $S_{r}(K_{S})\sim$46\% (last obs.). Figure \ref{images} shows the two K$_{S}$ images obtained during these two nights.

During the last observation, a PSF star was also acquired in order to perform accurate photometry of the adaptive optics data of the corresponding night. The object, \object{DENIS-P~J131500.9-251302} (spectral type $\sim$M8, J=15.2, H=14.54 and K$_{S}$=14.02~mag),  was observed under better conditions ($\lambda/r_{0}$=0\farcs43 and airmass=1.03) with a strehl ratio of $S_{r}(H)\sim$10\%, and $S_{r}(K_{S})\sim$40\%. Unfortunately it was not observed in J. We performed the photometry using standard DAOPHOT PSF fitting photometry. The results are summarized in Table \ref{rel_photom}. Although the PSF star has a spectral type earlier than 2MASSW~J0746425+2000321 and was observed at much better airmass and better seeing, the relative photometry we obtain in H and K$_{S}$ is in very good agreement with the one reported by \citet{2003ApJ...587..407C} with Gemini North/Hokupa'a and the one we measure with Keck I/NIRC (K$_{S}$ band).

\subsection{Speckle Observations with Keck}
On 2003 December 04, we obtained $K$ band speckle observations of our target at the 10~m Keck~I telescope with the facility instrument NIRC \citep{1994ExA.....3...65K}. With its re-imaging optics \citep{1996PASP..108..615M}, this 256$\times$256 near-infrared array offers a 0\farcs0203$\pm$0\farcs0003 pixel scale that provides Nyquist sampling of the diffraction limit of the telescope at this wavelength (about 0\farcs05); its absolute orientation is known to within 1\degr. Several stacks of 200 short-integration exposures were obtained ($t\sim0.1$~s, i.e., fast enough to effectively ``freeze'' the atmospheric turbulence and retain the high-angular resolution information in the image), and similar stacks on two calibration point sources were obtained immediately before and after our target. Standard speckle data reduction routines were applied to the data; we refer the reader to \citet{1993PhDT.........1G} and \citet{1998AJ....115.1972P} for more details and only summarize briefly the various stages involved in the data reduction process.  Each individual exposure is first sky subtracted, flat-fielded and bad pixel-corrected; its power spectrum is then calculated. The power spectra are median-averaged over each stack and divided by that of the calibrator. A 2-D sinusoidal function is then fitted to the power spectrum to determine the binary properties: separation, position angle and flux ratios. Uncertainties are estimated from the standard deviation of the parameters extracted from all stacks. There is a 180\degr\ ambiguity in the position angle of the binary as derived through power spectrum analysis, but this can be resolved by shift-and-adding all individual exposures using the brightest speckle  as a reference. The resulting image shows the companion to be roughly to the Northeast of the primary and the astrometric accuracy of the power spectrum analysis is much higher. The results obtained are reported in Tables \ref{rel_photom} and \ref{rel_astro} and Figure \ref{images} shows the final image. As mentioned above, the difference of magnitude is in very good agreement with the previous measurements within the uncertainties.

\subsection{High Angular Resolution -- Low Spectral Resolution Optical Spectroscopy with STIS \label{stis_analysis}}

In order to get the spatially resolved spectra of each component of the multiple system, we tried to align the slit along the axis of the binary. Scheduling constraints of HST made it difficult to get long slit STIS observations of 2MASSW~J0746425+2000321 at a particular roll angle of HST. In order to ease scheduling, a range of admitable roll angle was defined. This combined with the relatively rapid orbital motion of 2MASSW~J0746425+2000321 meant that the entrance slit of STIS was not optimally aligned with the position angle of 2MASSW~J0746425+2000321 at the time of the observations. Since the size of the slit we used (0\farcs2) is larger than the separation of the binary ($\sim$0\farcs125), we could nevertheless obtain a resolved 2-D spectrum and perform an extraction of the two spectra. The effects on the spectral analysis can be the following: since the red and the blue photocentres of each component are not symmetrically centred in the slit, the dispersion of the red and blue parts of the spectrum suffers differently from obstruction by the slit. Since the position of the photocentres and the dispersion of the light depends on the wavelength, the loss in flux also varies with the wavelength. This effect produces a ``bluer'' spectrum for the secondary.

\begin{figure*}
   \centering
   \includegraphics[width=\textwidth]{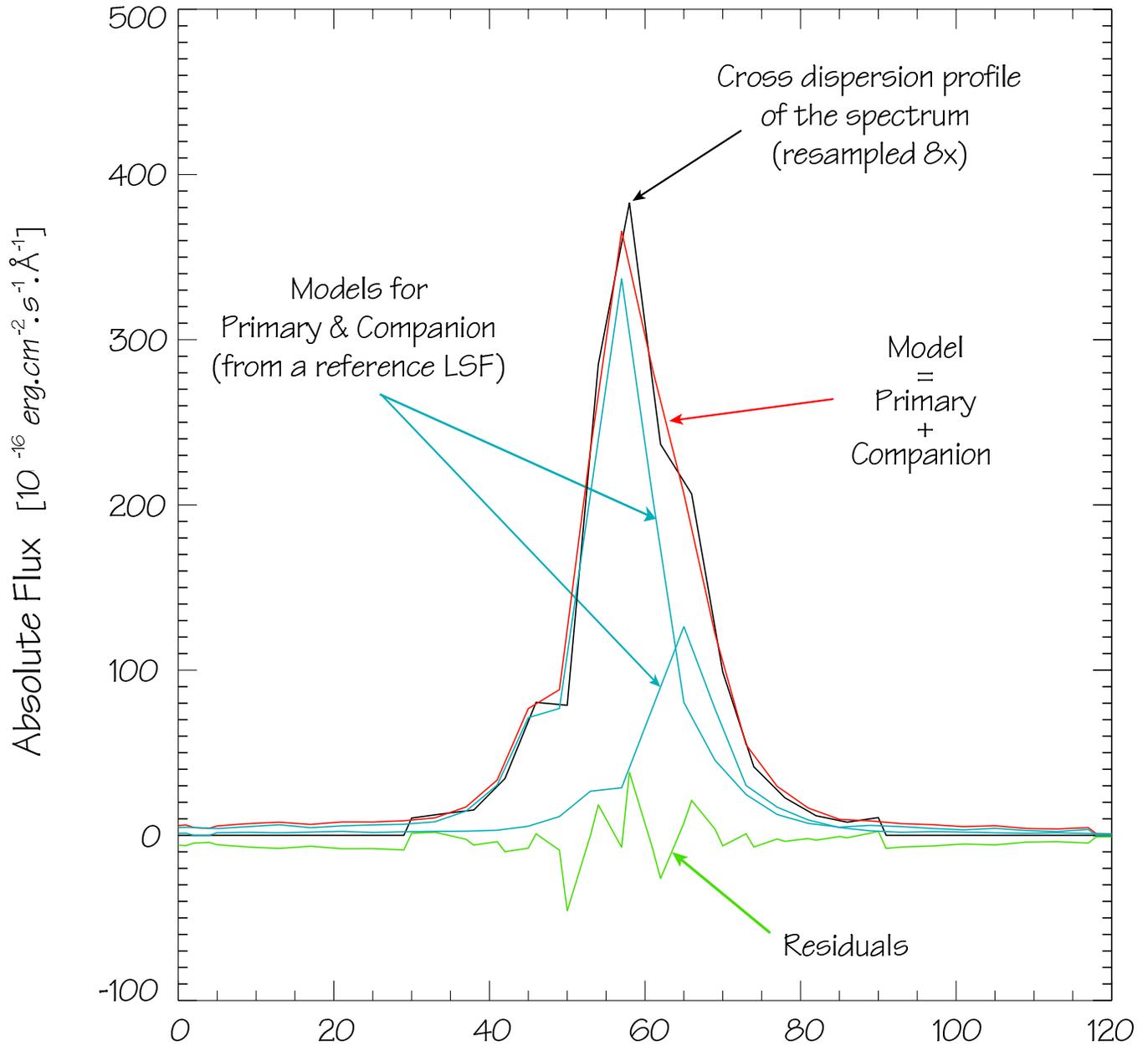}
   \caption{Extraction of the individual spectra. This figure shows the cross dispersion profile of the 2-D spectrum around 850~nm, and the best fits of the primary and the secondary (light blue). The sum of the primary and secondary (red) is also indicated for comparison with the raw data (black). In that case the intensity of the residuals (green) is less than $\sim$2\% of the intensity of the raw data. \label{lsf}}
   \end{figure*}

The separation between the two spectra is about 2 pixels, whereas the full width at half maximum (FWHM) of the line spread function (LSF) varies between 1.0 and 1.2 pixel, so that the two spectra are barely resolved. In order to extract the spectrum of each component, we used a custom made program able to perform a fit of the two blended spectra. On each cross-dispersion column, a minimum $\chi^{2}$ fit was performed to the data using the cross-dispersion profile of a reference spectrum at the same wavelength. The latter spectrum was obtained with the same instrument settings on a K7 dwarf (\object{TWA 6}, program 8176, P.I. Schneider). The free parameters for the fit are the amplitude of the primary, the amplitude of the secondary, the position of the primary and the position of the secondary. Since the cross dispersion profile is barely sampled, we also performed a linear re-sampling of the data by a factor of eight prior to the fit, in order to avoid problems due to spectral aliasing. To ensure more robustness and increase reliability, the program was used in 2 passes. A polynomial fit of the results on the positions of the two individual spectra was made after the first pass, in order to identify and remove outliers (due to bad pixels or cosmic rays). The results of these fits were then used as first guess inputs for the second pass. The results obtained with the second pass are very close to that obtained with the first pass but cleaner (without the more obvious bad pixels, cosmic rays and outliers), ensuring that the whole algorithm is robust enough and converging properly. Figure \ref{lsf} gives an overview of the results at a particular wavelength. The residuals after the fit represent between 1.5\% and 9\% of the total intensity of the original spectrum depending on the wavelength, thus of the same order than the signal to noise ratio, ensuring that the quality of the fit is good.

\section{Orbital parameters and determination of the total mass \label{analysis}}

We used three different and independent custom-made programs to determine the best fitting orbital solution for 2MASSW~J0746425+2000321 and the uncertainties on each of the fitted parameters. The orbit can be entirely described by seven independent parameters: semi-major axis ($a$), orbital period ($P$), inclination ($i$), eccentricity ($e$), position angle of the ascending node ($\Omega$), angle between the ascending node and periastron ($\omega$) and time of periastron passage ($T_0$). With seven two-dimension astrometric data-points, this fully-constrained problem has seven free parameters. The total mass of the system can be derived from the orbital period and semi-major axis through Kepler's Third Law.

\subsection{``\emph{Amoeba}'' method}

The first method minimizes in the nonlinear 7-dimensions function by downhill simplex method, using the \emph{amoeba} algorithm \citep[see e.g][ for a description of the method and algorithm.]{Numerical...Recipes}. It fits all seven orbital parameters simultaneously, taking into account non-equal errors of the measurements. The {\it reduced-}$\chi^{2}$ of 1.41 ensures that the fit is satisfactory (see section \ref{chi2} below). The results are shown in Figure \ref{orbit} and Table \ref{orbital_param}. No uncertainties on the derived parameters are available with this method. 

\subsection{Iterative method}

This method uses 50\,000 independent starting points that consist of a set of the 7 parameters being randomly chosen from their entire range of possible values. For each starting point, a Powell convergence algorithm minimizing the total $\chi^2$ \citep{Numerical...Recipes} modifies simultaneously all 7 parameters until it converges to a local minimum. Once convergence for all 50\,000 sets of initial guesses has been achieved, we read through the output file to find the absolute minimum of the $\chi^2$ function, which reveals the best-fitting orbital solution. Our best-fitting solution, illustrated in Figure \ref{orbit} has a satisfying {\it reduced-}$\chi^2$ value of 1.38.

\begin{figure*}
   \centering
   \includegraphics[width=\textwidth]{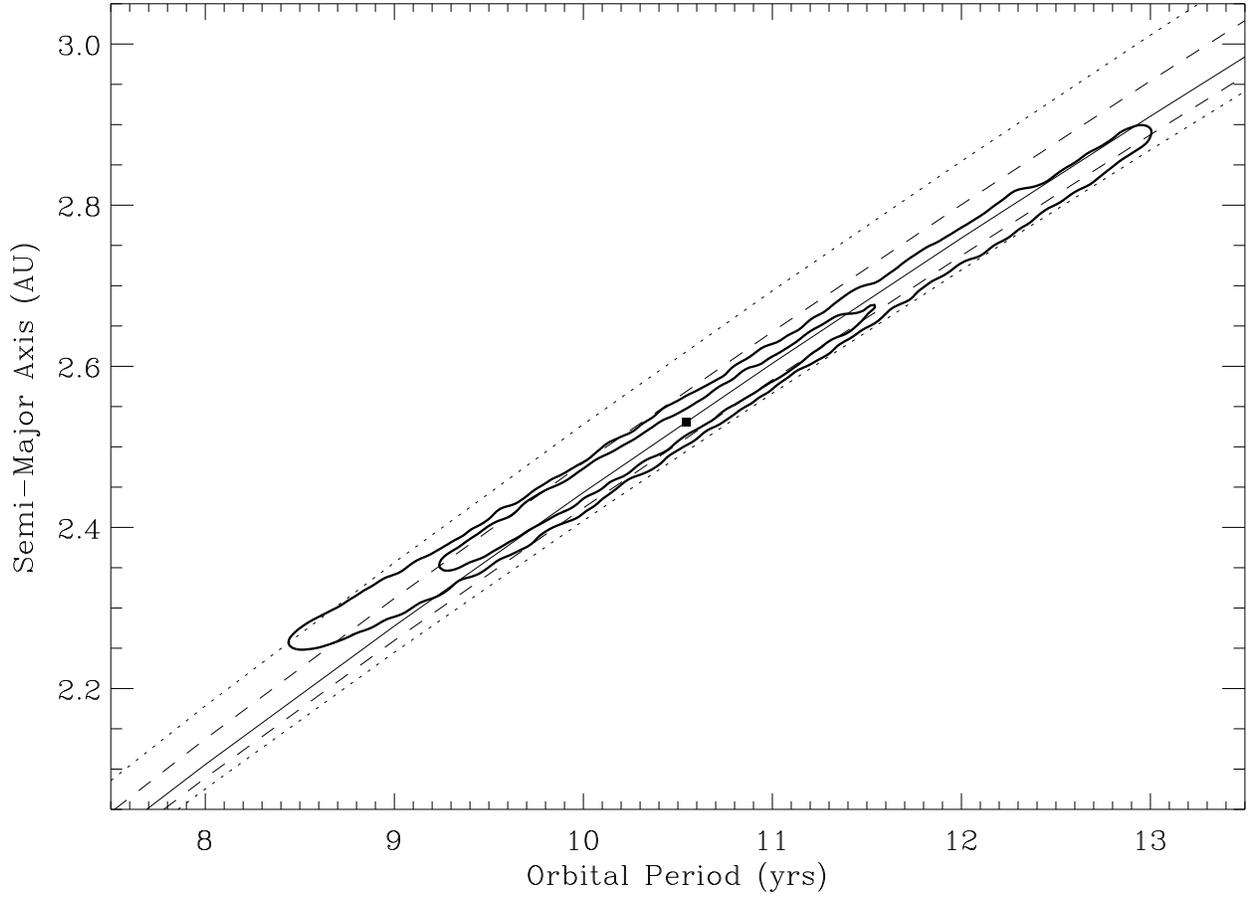}
   \caption{Ambiguity in the orbital parameters: the contours represent the solutions at 1- and 2-$\sigma$ ($\Delta\chi^{2}$=1.0 and 4.0 respectively) in the semi-major axis vs period space. The lines corresponding to different masses indicate the levels of confidence at 68.7\% and 95.4\% (1- and 2-$\sigma$ respectively). The filled square in the centre corresponds to the best fit. Although relatively large ranges are possible for the period and the semi-major axis, the range of corresponding acceptable masses is very narrow: between 0.143 and 0.153~M$_{\sun}$ at 1-$\sigma$ (represented by dashed lines) and between 0.141 and 0.160~M$_{\sun}$ at 2-$\sigma$ (dotted lines). Although the orbit is not perfectly known yet, the mass is relatively precisely determined, with a best value at 0.146$_{-0.006}^{+0.016}$~M$_{\sun}$ (2-$\sigma$ uncertainties, solid line).\label{sol_p_a}}
   \end{figure*}

Uncertainties for each parameters are defined by the range of possible values indicated by all solutions with {\it total} $\chi^2$ between $\chi^2_{min}$ and $\chi^2_{min} + 4$. These represent the 95.4\% confidence level for each parameters. Due to the highly non-linear behavior of the equations of orbital motion, the uncertainties do not follow a Gaussian statistics and nor even symmetric about the best fit. Note that the uncertainties are derived under the assumption that all parameters are independent, which is not correct. For instance, the uncertainties derived for $P$ and $a$ would yield an uncertainty on the system mass on order of $\pm0.055\,M_\odot$, $\sim$4 to 9~times larger than we actually derived here. Therefore, the uncertainties quoted here are only valid if they are used for one parameter at a time. Figure \ref{sol_p_a} shows that this is because the fitted values of $P$ and $a$ are tightly correlated and correspond to a very narrow range of possible masses. Although the orbit is not perfectly known yet, the total mass is relatively precisely determined: M$_{tot}$=0.146$^{+0.016}_{-0.006}$~M$_{\sun}$, corresponding to 4$\sim$11\% uncertainty.

The uncertainty on the distance to the system translates into a separate 2.4\% uncertainty (2-$\sigma$), or 0.035~M$_{\sun}$, on the system mass. The uncertainty on the orbital fitting is therefore the major source of uncertainty for this binary system since both sources of uncertainty (fit and distance) should be added in quadrature.

\subsection{\emph{ORBIT}}
We also used the \emph{ORBIT} program of \citet{1999A&A...351..619F}, fully described in their article. Briefly, ``{\it the program performs a least square adjustment to all available observations, with weights inversely proportional to the square of their standard errors [\ldots] The program uses a Levenberg-Marquardt minimization algorithm \citep{Levenberg...Marquadt} [...] Standard errors for derived parameters are computed from the full covariance matrix of the linearized least square adjustment.}'' 

\subsection{Reduced-$\chi^{2}$ and uncertainties \label{chi2}}

The reduced-$\chi^{2}$ values of $\sim$1.4 indicate that some of the uncertainties on the astrometric measurements may be slightly underestimated, although with only seven measurements for seven free parameters, such a value is statistically acceptable. Although rescaling the astrometric uncertainties to reach a reduced-$\chi^2$ of 1.0 could be argued for, the diversity of the instruments used for this orbital analysis suggests that such a treatment would be at least as erroneous as it could be helpful. For the time being, we decided to stick to the quoted astrometric uncertainties to derive the uncertainties on the orbital parameters of the binary.

\section{Discussion \label{discussion}}

\subsection{Spectral Types, Effective Temperatures \label{spt}}
The composite spectrum of 2MASSW\-~J0746425\-+2000321~AB in the optical has been previously studied by \citet{2000AJ....120..447K}, who derived a spectral type of L0.5. 

\begin{figure*}
   \centering
   \includegraphics[width=\textwidth]{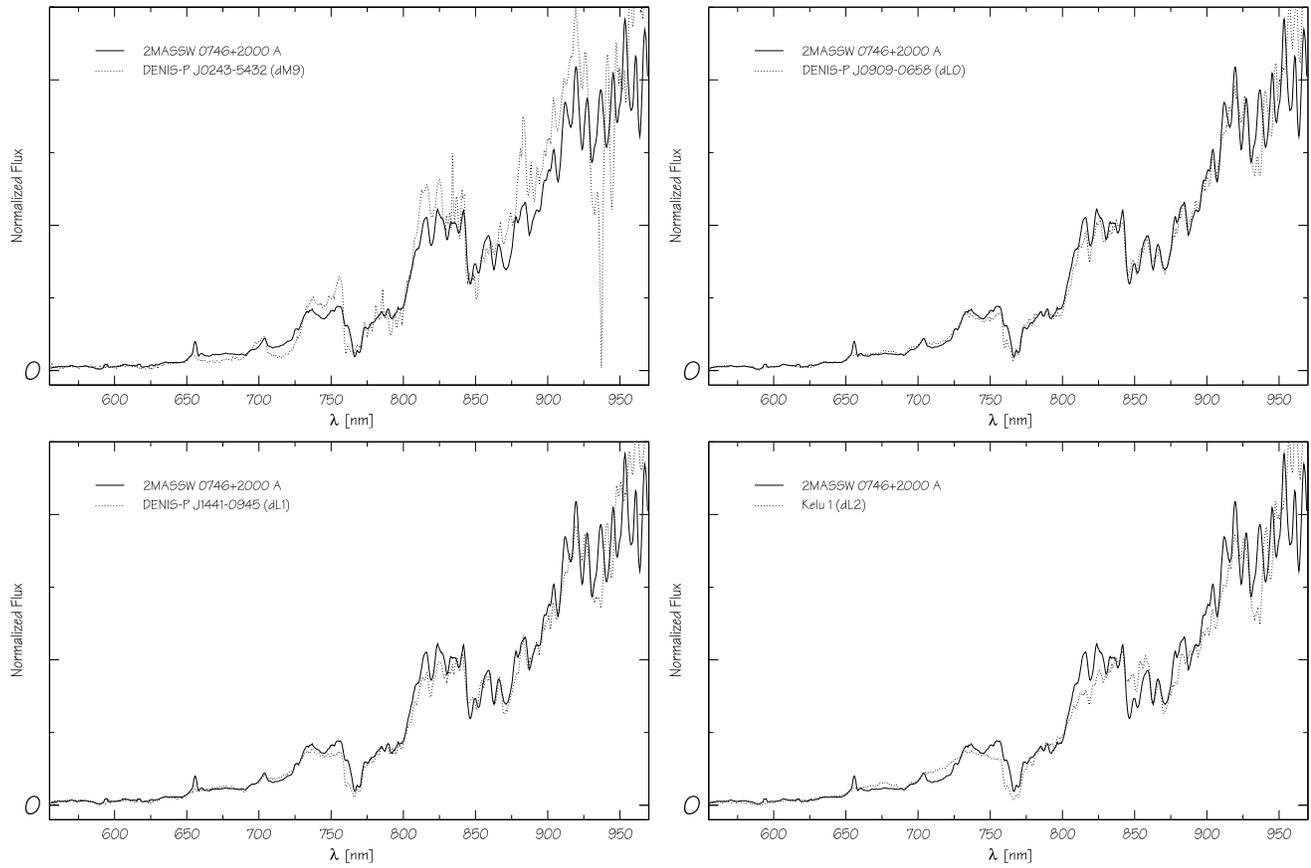}
   \caption{STIS optical low resolution spectra of 2MASSW~J0746425+2000321A compared to spectra of filed ultra-cool dwarfs. The four plots show the STIS spectrum of 2MASSW~J0746425+2000321A, smoothed via a boxcar (width = 5 pixels), and compared to: a) \object{DENIS-P~J024351.0-543219} (dM9); b) \object{DENIS-P~J090957.1-065806} (dL0); c)  \object{DENIS-P~J144137.3-094559} (dL1); d)  \object{Kelu 1} (dL2). All spectra have been normalized at 840~nm. Spectra for the field dwarfs from \citet{1999AJ....118.2466M}. \label{comp_spt_prim}}
   \end{figure*}

\begin{figure*}
   \centering
   \includegraphics[width=\textwidth]{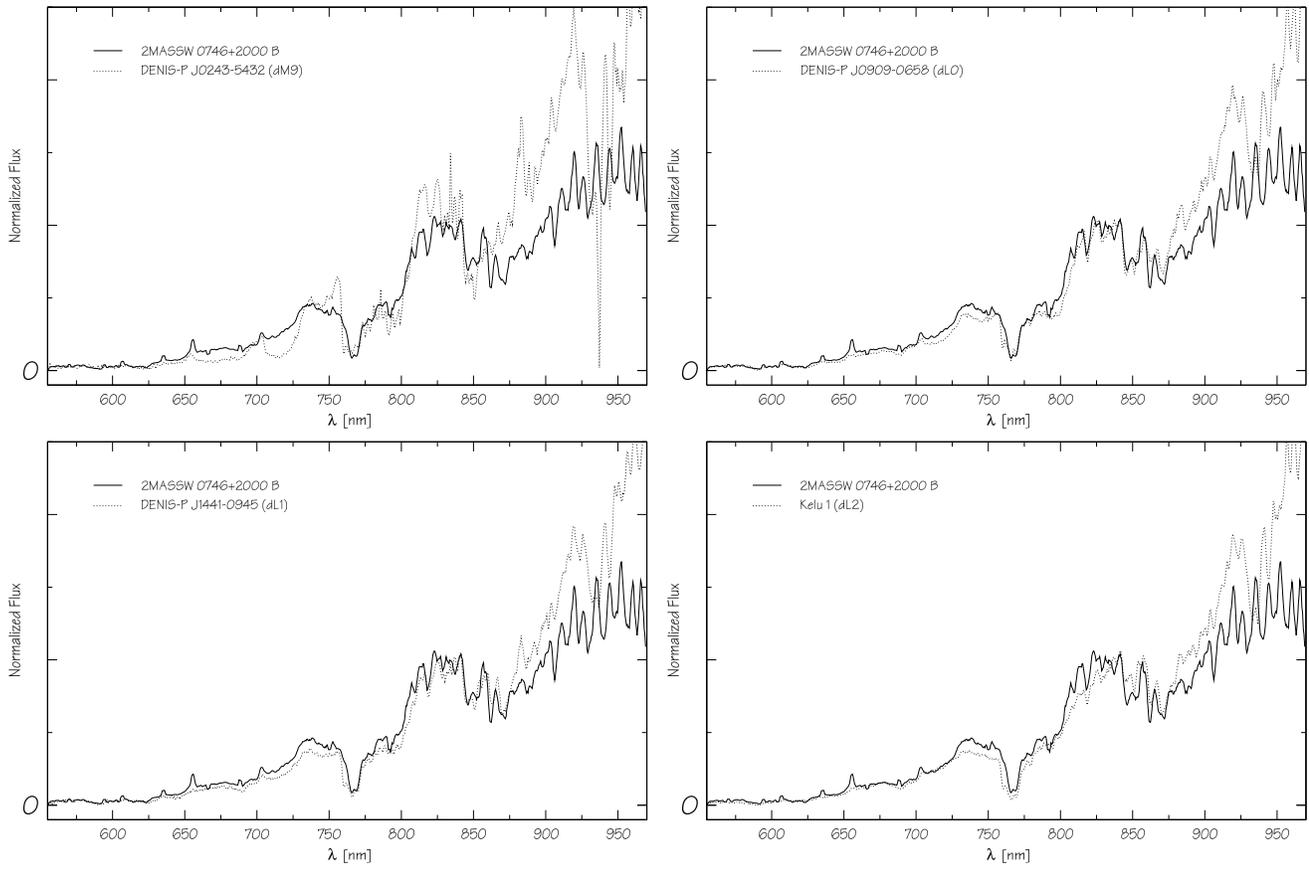}
   \caption{Same as Figure \ref{comp_spt_prim} but for 2MASSW~J0746425+2000321B \label{comp_spt_sec}}
   \end{figure*}

In order to derive the spectral types of each component, we compared their STIS spectra, extracted with the procedure described in Section \ref{stis_analysis}, with the spectra of field L dwarfs published by \citet{1999AJ....118.2466M}. As explained in this latter article, the relative strength of the TiO bands between 840--860~nm and the CrH and FeH bands between 860--880~nm are good indicators for the effective temperature changes. In later L dwarfs the TiO bands get weaker with respect to the CrH and FeH bands. Figures \ref{comp_spt_prim} and \ref{comp_spt_sec} shows the comparison. 2MASSW\-~J0746425\-+2000321~A is clearly very similar to the L0 field dwarf DENIS-P~J090957.1-065806, while 2MASSW\-~J0746425\-+2000321~B is between the L1 (DENIS-P~J144137.3-094559) and L2 (Kelu 1) field dwarfs. We thus derive a spectral type of L0$\pm$0.5 and L1.5$\pm$0.5 for A and B, respectively. This is consistent with the spectral type obtained by \citet{2000AJ....120..447K}, which is a blend of A and B. It is also consistent with the modest difference in brightness (see Table \ref{rel_photom}), which implies a difference in temperature of only 100 K according to the models of \citet{2000ApJ...542..464C}.

\subsection{Spectral Features}

The main emission lines present in the primary's spectrum is H$\alpha$ (EW=-25.0$\pm$0.5~\AA, 1-$\sigma$). The spectrum of the secondary is more noisy but the H$\alpha$ emission line appears clearly, with an equivalent width of EW=-18.0$\pm$0.5~\AA. From their high-resolution spectra, \citet{2002AJ....124..519R} reported a H$\alpha$ emission of -1.2~\AA~ for the unresolved system. The difference between the two measurements indicates that 2MASSW~J0746425+2000321 A and B display some chromospheric and/or magnetic activity. Li~\textsc{i} absorption is not detected in any of the two components with an upper limit of $\sim$1.5~\AA. \citet{2002AJ....124..519R} did not detect any Li~\textsc{i} absorption with an upper limit of detection at $\le$0.5~\AA. The presence of strong resonance doublets of alkali elements (K~\textsc{i} at 766.5 and 769,9~nm; Na~\textsc{i} at 818.4 and 819.5~nm; and Cs~\textsc{i} at 852.1 and 894.3~nm) as well as strong metallic molecular band-heads (CrH at 861.1~nm and FeH at 869.2~nm) is characteristic for L-dwarfs \citep{1997A&A...327L..29M}. The measurements of equivalent widths of the main atomic lines are reported in Table \ref{atomic_lines}. It is interesting to note that the equivalent width we measure for Cs~\textsc{i} at 852.1~nm for the primary corresponds to an effective temperature of $\sim$1900--2000~K in the effective temperature scale of \citet{2000ApJ...538..363B}, therefore in good agreement with the effective temperature derived by \citet{2001ApJ...555..368S} from their comparison of low and high resolution Keck spectra with the DUSTY models.

\subsection{Colour-Magnitude Diagrams \label{hr}}

\begin{figure*}
   \centering
   \includegraphics[width=\textwidth]{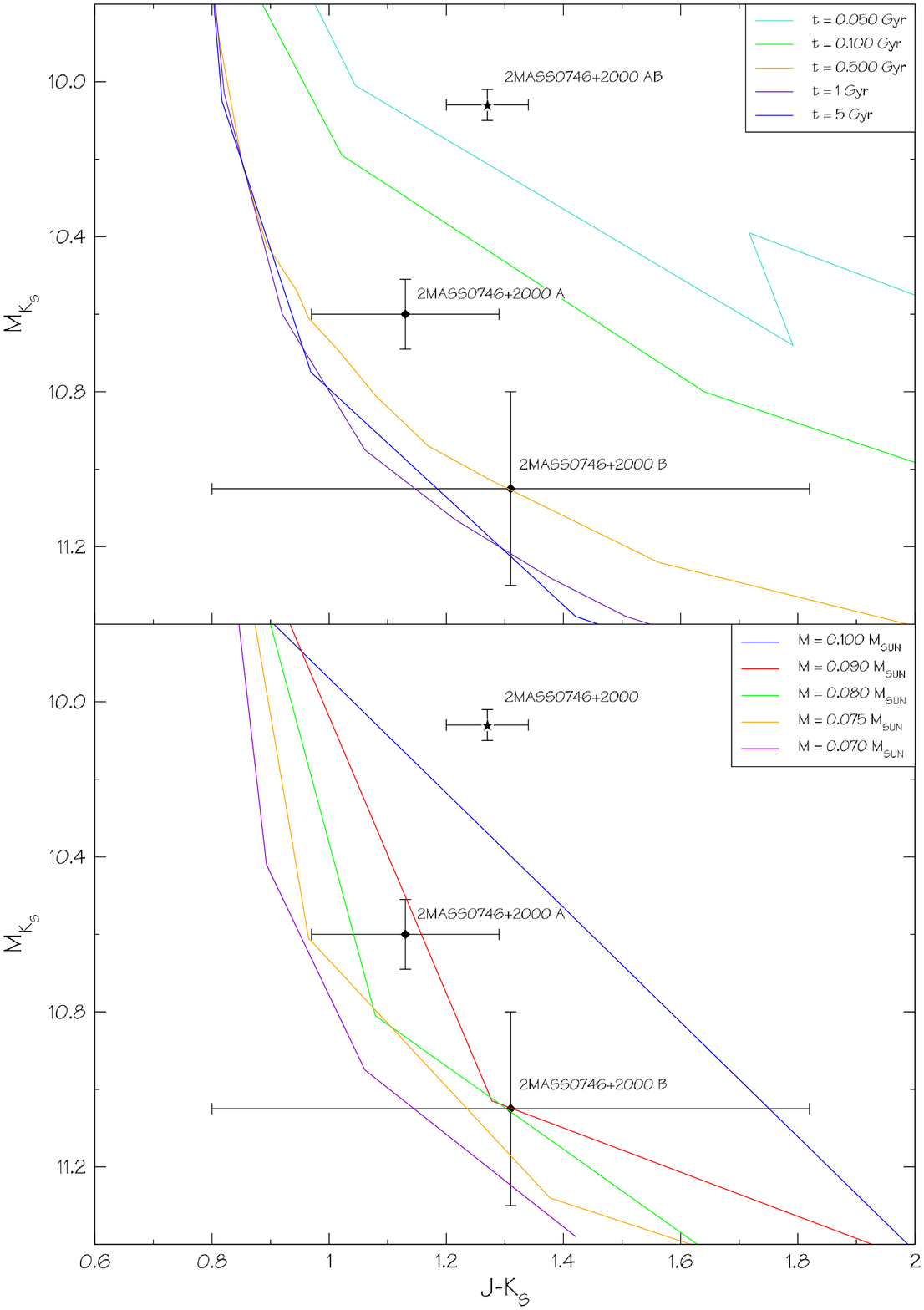}
   \caption{Colour-Magnitude diagrams M$_{K_{S}}$ vs ($J-K_{S}$) displaying the location of 2MASSW~J0746425+2000321A, B and AB (combined light). The 1--$\sigma$ combined uncertainties include the uncertainty on the distance. Isochrones of the DUSTY models \citep{2000ApJ...542..464C} are over-plotted for different ages (upper panel) and different masses (lower panel). \label{k_jk}}
   \end{figure*}

Figure \ref{k_jk} shows a colour-magnitude diagram of 2MASSW\-~J0746425+2000321~AB, 2MASSW\-~J0746425+2000321~A and 2MASSW\-~J0746425+2000321~B, and compares with the isochrones of the most recent DUSTY models for solar metallicity. To convert the observed magnitudes to absolute magnitudes we used the trigonometric parallax reported by \citet{2002AJ....124.1170D}. 

The position of 2MASSW~J0746425+2000321A shows that the age ranges between 150 and 500~Myr, thus relatively young. This is not consistent with the very high surface gravity obtained by \citet{2001ApJ...555..368S}. They used high and low resolution unresolved spectra and compared it with the DUSTY atmospheric models of \citet{2001ApJ...556..357A}. Using a $\chi^{2}$-fitting algorithm, they obtain an effective temperature of 2000~K and a surface gravity  log~$g$=6.0 from their low resolution spectra. From their high resolution spectra, they obtain an effective temperature of 1900$\sim$2000~K and a surface gravity  log~$g$=5.0$\sim$5.5. The temperatures are in good agreement with the spectral types  and colours we report here, but the surface gravity is too high for the young age. This is probably because the DUSTY models overestimate the dust effects. As a consequence, the strength of the alkali lines in the optical decreases, and the surface gravity is biased toward higher values.

The mass of the primary ranges between 0.075 and 0.095~M$_{\sun}$, while the mass of the secondary has large error bars and ranges between 0.055 and 0.100~M$_{\sun}$. The total mass of the system therefore ranges between $\sim$0.130 and $\sim$0.190~M$_{\sun}$, which is consistent with the dynamical mass considering the large uncertainties in the H-R diagram. 

For the secondary's mass it appears more appropriate to use the mass of the primary from the H-R diagram, which has reasonable uncertainties, together with the very precise dynamical total mass. This yields a mass between 0.052 and 0.072~M$_{\sun}$, therefore clearly substellar. The absence of lithium absorption in the spectra gives also a constrain on the lower limit of the mass. According to the DUSTY evolutionary models, Lithium should be depleted for masses greater than 0.075~M$_{\sun}$ at 150~Myr and masses greater than 0.060~M$_{\sun}$ at 500~Myr. The mass of the secondary must  therefore be greater than 0.060, and ranges between 0.060$\le$M$_{B}\le$0.072~M$_{\sun}$.

The system is thus very likely made of a brown dwarf orbiting a slightly more massive very low mass star. Both objects are very close to the stellar-substellar boundary.

\section{Conclusion}

We present new astrometric, photometric and spectroscopic observations of the ultra-cool 2MASSW\-~J0746425\-+2000321AB pair, which have enabled us to compute the orbital parameters, total and individual masses and spectral types of the system. We find a total mass of 0.146$^{+0.016}_{-0.006}$~M$_{\sun}$ (2-$\sigma$), with a L0$\pm$0.5 primary and a L1.5$\pm$0.5 secondary. The orbit is eccentric ($e=0.41^{+0.08}_{-0.09}$) with a period of 3850.9$^{+904}_{-767}$~days. Our observations enable us to follow 60\% of the orbit, corresponding to 36\% of the period. Near-infrared photometry of the individual components enables us to locate them in a H-R diagram and compare them to the most recent evolutionary DUSTY models. The pair is likely formed by a brown dwarf (L1.5$\pm$0.5, with 0.060$\le$M$_{B}\le$0.072~M$_{\sun}$) orbiting a L0$\pm$0.5 very low mass star (0.075$\le$M$\le$0.095~M$_{\sun}$). The system appears to be young, with an age in the range 150$\sim$500~Myr. 
Further high-angular resolution images should allow to cover 100\% of the orbit and refine the orbital parameters. Radial velocity measurements, if feasible, would then allow to compute precise individual masses.

\begin{acknowledgements}
     The authors thank the referee of this manuscript, Dr M. Simon for the useful comments and advices he made on this paper. This work is based on observations collected at the European Southern Observatory (Paranal, Chile), program 70.D-0773; with the NASA/ESA Hubble Space Telescope obtained at the Space Telescope Science Institute (STSci), programs GO8720 and GO9451, and at the Keck Observatory (Hawai'i, U.S.A). We would like to acknowledge the great cultural significance of Mauna Kea for native Hawaiians, and express our gratitude for permission to observe from its summit. The STSci is operated by the Association of Universities for Research in Astronomy, Inc., under NASA contract NAS 5-26555. This publication makes use of data products from the Two Micron All Sky Survey, which is a joint project of the University of Massachusetts and the Infrared Processing and Analysis Center/California Institute of Technology, funded by the National Aeronautics and Space Administration and the National Science Foundation. This work made use of data from the The Guide Star Catalogue~II, which is a joint project of the Space Telescope Science Institute and the Osservatorio Astronomico di Torino. The participation of the Osservatorio Astronomico di Torino is supported by the Italian Council for Research in Astronomy. Additional support is provided by European Southern Observatory, Space Telescope European Coordinating Facility, the International GEMINI project and the European Space Agency Astrophysics Division.
\end{acknowledgements}


\newpage

\renewcommand{\arraystretch}{1.5}


   \begin{table*}
\caption[]{Astrometry and photometry from different catalogues}
\label{catalogues}
\footnotesize
\begin{tabular}{ lcccccll }
\hline
\hline
Date            &  R.A          &      Dec.     &  Uncert.         &   Filter & Mag.   &  Source  & Ident.  \\
\textsc{dd/mm/yyyy}   &  (J2000)      &   (J2000)     &                  &          & [mag]  &          &      \\
            \hline
01/01/1984\footnotemark[1]      & 07 46 42.5    & +20 00 32.6 &  $\pm$0\farcs1     &    R1    &  18.28  &  USNO-B1.0 & \object{USNO-B1.0~1100-0150847} \\
                &               &               &                  &    B2  &  21.7  & \\
                &               &               &                  &    R2  &  17.87  & \\
\hline
01/01/1998      &  07 46 42.55 &  +20 00 32.14 &  $\pm$0\farcs3   &    R   &  17.6         &  GSC2.21  & \object{GSC 2W~22110125398}  \\
\hline
05/12/1997      &  07 46 42.56  &  +20 00 32.2  &  $\pm$0\farcs1   &    J   &  11.759  &  2MASS  &  \object{2MASSW~J07464256+2000321} \\
                &               &               &                  &    H   &  11.007 & &  \\
                &               &               &                  &    K   &  10.468 & &  \\
\hline
06/12/2002      &               &               &                  &    L'  &  11.19  & \citet{2002ApJ...564..452L} &  \\
     \hline
\end{tabular}

\thanks{\footnotemark[1] Mean epoch of observation}
\end{table*}


   \begin{table*}
	      \caption[]{Observation log.} 
         \label{obs_log}
     \begin{tabular}{lcccc}
            \hline
	    \hline
	     \multicolumn{5}{c}{Imaging} \\
	    \hline
            Instrument           &  Filter    & Exp.     & Date Obs.    & Pixel Scale \\
	                         &            & Time [s] & \textsc{dd/mm/yyyy} & [\arcsec]  \\
            \hline
	    HST/WFPC2-PC         & F814W      & 50       & 25/04/2000   & 0\farcs0455  \\
	    Gemini North/Hokupa  &   J        & 120      & 07/02/2002   & 0\farcs0199  \\
	    Gemini North/Hokupa  &   H        & 720      & 07/02/2002   & 0\farcs0199  \\
	    Gemini North/Hokupa  &   K        & 120      & 07/02/2002   & 0\farcs0199  \\
	    HST/ACS-HRC          & F625W      & 960      & 21/10/2002   & 0\farcs0250\footnotemark[1]  \\
	    HST/ACS-HRC          & F775W      & 440      & 21/10/2002   & 0\farcs0250\footnotemark[1]  \\
	    HST/ACS-HRC          & F850LP     & 340      & 21/10/2002   & 0\farcs0250\footnotemark[1]  \\
	    VLT/NACO             & K$_{S}$    & 0.4      & 18/02/2003   & 0\farcs0148\footnotemark[2]  \\
	    VLT/NACO             & J          & 10       & 22/03/2003   & 0\farcs0148\footnotemark[2]  \\
	    VLT/NACO             & H          & 5        & 22/03/2003   & 0\farcs0148\footnotemark[2]  \\
	    VLT/NACO             & K$_{S}$    & 5        & 22/03/2003   & 0\farcs0148\footnotemark[2]  \\
	    Keck I/NIRC          & K$_{S}$    & 20       & 04/12/2003   & 0\farcs0203  \\
	    HST/STIS             & longpass   & 10       & 09/01/2004   & 0\farcs0508  \\
            \hline
            \hline
	     \multicolumn{5}{c}{Spectroscopy} \\
	    \hline
            Instrument     &  Wavelength      & Exp.     & Date Obs.    & Dispersion  \\
	                   &  Range [nm]      & Time [s] & \textsc{dd/mm/yyyy} & [\AA/pixel]  \\
            \hline
	    HST/STIS       &  525--1300       & 1980     & 09/01/2004   & 4.92  \\
            \hline
     \end{tabular}

\thanks{\footnotemark[1] Effective value on the processed images, slightly different from the 0\farcs028$\times$0\farcs025 given in the manual.\\}
\thanks{\footnotemark[2] Effective value on the processed images, slightly different from the 0\farcs01326 given in the manual.}

   \end{table*}   


\begin{table*}
\caption[]{Relative Photometry of 2MASSW~J0746425+2000321AB}
\label{rel_photom}
\begin{tabular}{ lccccc }
\hline
\hline
Date       &  Instrument            &  Filter     &  Mag. Prim.   &  $\Delta$Mag. &  Source  \\
\textsc{dd/mm/yyyy} &                        &             &   [mag]       &    [mag]      &          \\
       \hline
25/04/2000 & HST/WFPC2              &  F814W      &  15.41$\pm$0.15  & 1.00$\pm$0.09 &   (1)     \\
07/02/2002 & Gemini North/Hokupa'a  &  J          &  12.19$\pm$0.07  & 0.60$\pm$0.20 &   (2)     \\  
07/02/2002 & Gemini North/Hokupa'a  &  H          &  11.54$\pm$0.11  & 0.48$\pm$0.15 &   (2)     \\  
07/02/2002 & Gemini North/Hokupa'a  &  K'         &  11.05$\pm$0.09  & 0.44$\pm$0.15 &   (2)     \\  
21/10/2002 & HST/ACS                &  F625W      &  18.81$\pm$0.05  & 0.48$\pm$0.03 &   (3)     \\  
21/10/2002 & HST/ACS                &  F775W      &  15.98$\pm$0.05  & 0.68$\pm$0.04 &   (3)     \\  
21/10/2002 & HST/ACS                &  F850LP     &  14.24$\pm$0.05  & 0.76$\pm$0.04 &   (3)     \\  
22/03/2003 & VLT/NACO               &  H          &  11.55$\pm$0.08  & 0.46$\pm$0.15 &   (3)     \\  
22/03/2003 & VLT/NACO               &  K$_{S}$    &  11.06$\pm$0.09  & 0.42$\pm$0.15 &   (3)     \\ 
04/12/2003 & Keck I/NIRC            &  K$_{S}$    &  11.03$\pm$0.03  & 0.52$\pm$0.03 &   (3)     \\ 
       \hline
\end{tabular}

\thanks{\footnotemark[1] Source: (1) \citet{2003AJ....126.1526B}; (2) \citet{2003ApJ...587..407C}; (3) this paper}
\end{table*}


\begin{table*}
\caption[]{Relative Astrometry of 2MASSW~J0746425+2000321AB \label{rel_astro}}
\begin{tabular}{ lccccc }
\hline
\hline
Date           &  Time\footnotemark[1] &     Sep.\footnotemark[2]       &  P.A\footnotemark[2] & Instrument  &  Source\footnotemark[4]  \\
\textsc{dd/mm/yyyy} &  \textsc{hh/mm/ss} &   [mas]        &  [\degr]      &                      &          \\
       \hline
25/04/2000     &   08:14:14     &  219$\pm$3     &  168.8$\pm$0.8   & HST/WFPC2                &  (1) \\
07/02/2002     &   09:48:55     &  121$\pm$8     &  85.7$\pm$3.5    & Gemini North/Hokupa'a     &  (2) \\
21/10/2002     &   23:10:43     &  119.5$\pm$1   &  33.9$\pm$0.5    & HST/ACS                  &  (3) \\
18/02/2003     &   01:40:45     &  131.3$\pm$3.9 &  13.8$\pm$1.9    & VLT/NACO                 &  (3) \\
22/03/2003     &   01:22:00     &  123.5$\pm$2.1 &  4.6$\pm$1.0    & VLT/NACO                 &  (3) \\
04/12/2003     &   15:15        &  126.5$\pm$1.8 &  317.9$\pm$0.7   & Keck I/NIRC              &  (3) \\
09/01/2004     &   18:51:45     &  134.5$\pm$3   &  311.1$\pm$1.2   & HST/STIS                 &  (3) \\
       \hline
\end{tabular}

\thanks{\footnotemark[1] The uncertainty corresponds to the exposure time (see Table \ref{obs_log})}\\
\thanks{\footnotemark[2] 1-$\sigma$ uncertainties (combined instrumental and measurement)}\\
\thanks{\footnotemark[3] Source: (1) \citet{2003AJ....126.1526B}; (2) \citet{2003ApJ...587..407C}; (3) this paper}
\end{table*}


\begin{table*}
\caption[]{Orbital Parameters of 2MASSW~J0746425+2000321AB}
\label{orbital_param}
\begin{tabular}{ lccc }
\hline
\hline
Parameter                                       &  Iterative Method\footnotemark[1]  &  \emph{Amoeba} Method  &  \emph{ORBIT}\footnotemark[2] \\
\hline
Total Mass [M$_{\sun}$]                         &    0.146$^{+0.016}_{-0.006}$       &   0.1511              &   0.148 \\
Period, $P$ [days]                              &    3850.9$^{+904}_{-767}$          &   3718                &   3863$\pm$609    \\
Eccentricity, $e$                               &    0.41$^{+0.08}_{-0.09}$          &   0.3999              &   0.417$\pm$0.062 \\
Semi-major axis, $a$, [A.U]                     &    2.53$^{+0.37}_{-0.28}$          &   2.50                &   2.55$\pm$0.25   \\
Inclination, $i$ [\degr]                        &    141.6$^{+2.5}_{-3.4}$           &   141.65              &   140.65$\pm$2.29 \\
Argument of Periapsis, $\omega$ [\degr]         &    350.6$^{+5.2}_{-5.9}$           &   350.15              &   350.65$\pm$3.58 \\
Longitude of ascending node, $\Omega$ [\degr]   &    20.7$^{+9.9}_{-14.2}$           &   18.97               &   20.84$\pm$7.68  \\
Periastron Passage, $T_{0}$ (year)              &    2002.89$^{+0.14}_{-0.09}$       &   2002.91             &   2002.84$\pm$0.07\\
reduced-$\chi^{2}$ of the fit                   &    1.38                            &   1.41                &   1.46 \\
\hline
\end{tabular}

\thanks{\footnotemark[1] 2-$\sigma$ uncertainties, corresponding to a 95.4\% level of confidence. These uncertainties do not include the uncertainty on the distance ($\sim$2.4\%, 2-$\sigma$), which should be added in quadrature.\\}
\thanks{\footnotemark[2] 2-$\sigma$ uncertainties. These uncertainties do include the uncertainty on the distance, but assume that the uncertainty are linear, which is not the case here.}
 \end{table*}


\begin{table*}
\caption[]{Atomic lines in the spectra of 2MASSW~J0746425+2000321A and B}
\label{atomic_lines}
\begin{tabular}{ lccccccc }
\hline
\hline
                         &  \multicolumn{2}{c}{K~\textsc{i} Wavelength}  & & \multicolumn{2}{c}{Cs~\textsc{i} Wavelength } & & Na~\textsc{i~d} Wavelength\footnotemark[1]\\
\cline{2-3} \cline{5-6} \cline{8-8} 
Object                   &  7665           &   7699           & &  8521            &      8943      & &  8183-8195  \\
\hline
2MASSW~J0746425+2000321A &  22.6           &   17.4           & &  2.1             &      0.99      & &  8          \\
2MASSW~J0746425+2000321B &  19.0           &   16.4           & &  \ldots          &      \ldots    & &  5          \\
\hline
\end{tabular}

\thanks{Note.--- All units are in angstr\"oms. 1-$\sigma$ uncertainties are $\sim$0.5~\AA.\\}
\thanks{\footnotemark[1] Corresponds to the blend of the 8183 and 8195~\AA\, doublet.}
\end{table*}


\newpage


   \begin{figure*}
   \centering
   \includegraphics[width=\textwidth]{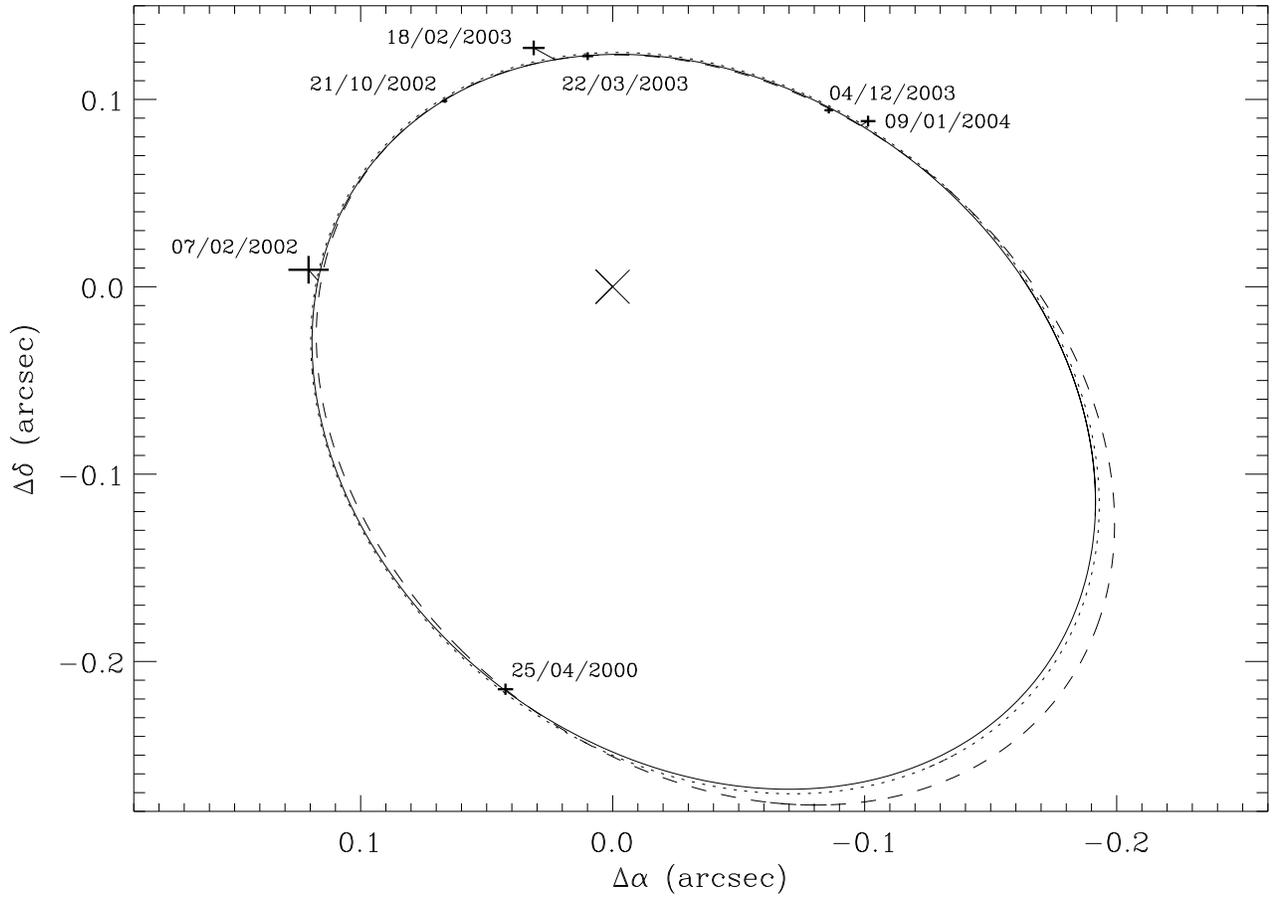}
   \caption{Positions of 2MASSW~J0746425+2000321A and B and best fit of the orbit \label{orbit}. The dotted curve represents the best bit orbit obtained with the \emph{amoeba} method, and the solid curve the result obtained with the iterative method, and the dashed curve the solution given by \emph{ORBIT}. It appears clearly that the three methods give close results, well within the uncertainties. The plus indicate the observations and their uncertainties, and the corresponding epoch is indicated. The central cross shows the position of the primary. }
   \end{figure*}


\bibliographystyle{aa}

\end{document}